\documentclass[prl,letterpaper,aps,10pt,superscriptaddress,twocolumn,floatfix,showpacs]{revtex4-1}
\usepackage{graphicx}
\usepackage{amsmath}
\usepackage{amsfonts}
\usepackage{amssymb}
\usepackage{epsfig}
\usepackage[pdftex]{color}
\usepackage{amsmath,graphicx,amssymb,braket,xcolor,subfigure,upgreek}
\usepackage[colorlinks, linkcolor=blue, citecolor=blue, urlcolor=blue, breaklinks=true]{hyperref}
\usepackage{microtype}
\usepackage{bbm}
\usepackage{color}

\newcommand{\fref}[1]{Fig.~\ref{#1}}
\newcommand{\ts}[1]{_\text{#1}}
\renewcommand{\section}[1]{\emph{#1} --}

\begin{document}

\title{Cavity antiresonance spectroscopy of dipole coupled subradiant arrays}
\author{David Plankensteiner}
\affiliation{Institut f\"ur Theoretische Physik, Universit\"at Innsbruck, Technikerstr. 21a, A-6020 Innsbruck, Austria}
\author{Christian Sommer}
\affiliation{Max Planck Institute for the Science of Light, Staudtstra{\ss}e 2,
D-91058 Erlangen, Germany}
\author{Helmut Ritsch}
\affiliation{Institut f\"ur Theoretische Physik, Universit\"at Innsbruck, Technikerstr. 21a, A-6020 Innsbruck, Austria}
\author{Claudiu Genes}
\affiliation{Max Planck Institute for the Science of Light, Staudtstra{\ss}e 2,
D-91058 Erlangen, Germany}
\date{August 31, 2017}

\begin{abstract}
An array of $N$ closely spaced dipole coupled quantum emitters exhibits super- and subradiance with characteristic tailorable spatial radiation patterns. Optimizing the emitter geometry and distance with respect to the spatial profile of a near resonant optical cavity mode allows us to increase the ratio between light scattering into the cavity mode and free space emission by several orders of magnitude. This leads to distinct scaling of the collective coherent emitter-field coupling versus the free space decay as a function of the emitter number. In particular, for subradiant states, the effective cooperativity increases much faster than the typical linear $\propto N$ scaling for independent emitters. This extraordinary collective enhancement is manifested both in the amplitude and the phase profile of narrow collective antiresonances appearing at the cavity output port in transmission spectroscopy.
\end{abstract}


\maketitle

The confinement of atoms and photons in small volumes with very low loss has been a renowned success~\cite{haroche1989cavity,walther2006cavity,kimble1998strong} as it allows for tests of light-matter interactions where the quantum nature of both comes into play. In a cavity quantum electrodynamics setup, the photon-emitter interaction strength $g_1 \propto  \mu \mathcal{E}$ for an emitter with a dipole moment $\mu$ is strongly enhanced by decreasing the field mode volume and, thus, increasing the local field per photon $\mathcal{E}$. In a standard Fabry-P\'{e}rot cavity geometry, this is achieved by closely surrounding the emitter with two high-reflectivity mirrors. The atom-photon interaction time is then enhanced by a factor roughly proportional to the cavity finesse characterizing the number of round trips a photon can make before escaping to the environment at a rate $\kappa$. At the single quantum emitter level, this has facilitated experimental progress towards strong coupling allowing the study of single photon nonlinear effects, such as the photon blockade regime~\cite{birnbaum2005photon}, of vacuum Rabi splittings and other tests of fundamental quantum optics effects~\cite{yoshie2004vacuum,casanova2010deep}.

The single emitter cooperativity $C_1 = g_1^2/(\kappa \gamma)$
(where $\gamma$ is the rate of spontaneous decay into free space) is a well established measure for strong light-matter interaction when $C_1\gg 1$. Since, for a single two-level emitter, the dipole matrix element $\mu$ appears both in $g_1\propto \mu$ and $\gamma \propto \mu^2$, the cooperativity $C_1$ is merely a geometric factor independent of $\mu$ \cite{vuletic2001three}. This means that cavity design (increasing the finesse and decreasing the transverse mode area) is the central aspect for reaching high single emitter cooperativity. In the parameter regime of large $\kappa$, one often targets a large effective cooperativity by coupling $N$ emitters simultaneously to the same cavity mode. For distant fully independent emitters, the effective cooperativity then scales like $C\ts{eff}= C_1 N$, as the emitter-cavity coupling $g_{N}=g_1 \sqrt{N}$ increases proportionally to $\sqrt{N}$, while the free space emission rate $\gamma$ stays constant.
However, especially for small emitter-emitter separations, their coupling to the vacuum modes is inherently \textit{collective} generating states with superradiant and subradiant decay \cite{plankensteiner2015selective}, which invalidates the above simple scaling law. Such decay processes have recently attracted interest in 1D and 2D subwavelength spaced atomic arrays used in topological quantum optics, high extinction media or photon storage~\cite{bettles2015cooperative, bettles2016cooperative, bettles2016enhanced, shahmoon2016cooperativity,perczel2017topological,asenjo2017exponential}.

\begin{figure}[b]
\includegraphics[width=0.9\columnwidth]{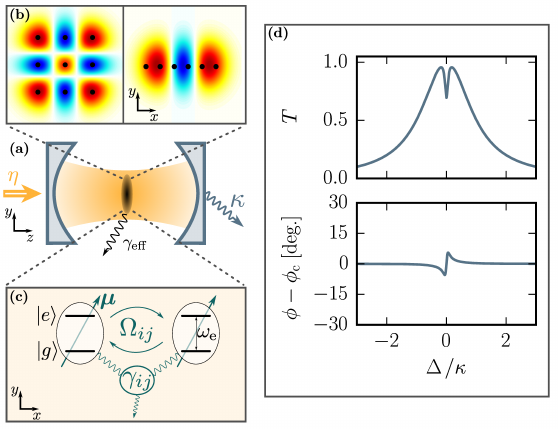}
\caption{\emph{System setup}. (a) Optical cavity supporting (b) different transverse modes coupled to (c) a rigid array of dipole-dipole interacting quantum emitters. (d) Light-matter interaction creates an antiresonance dip and a fast phase switch around the emitter resonance in the cavity transmission spectrum as shown on the right for a single emitter with $g_1=\kappa/10=2\gamma$.}
\label{fig1}
\end{figure}

We introduce an alternative, improved path, towards reaching a high cooperativity based on collective dissipative effects. The mechanism involves the separate optimization of the coherent coupling of the emitters to the cavity mode and of the incoherent emitter-vacuum coupling. For a configuration of $N$ closely spaced emitters (separation less than the transition wavelength $\lambda\ts{e}$), the coupling to free space vacuum modes can be strongly suppressed~\cite{dicke1954coherence,haroche1982superradiance,plankensteiner2015selective}. At the same time, a periodic arrangement of emitters in a rigid geometry (for example implanted inside a solid-state matrix), transversely placed inside a single cavity mode can lead to optimized collective coupling~\cite{zoubi2008bright}. The upshot is that $C\ts{eff}$ scales strongly in a nonlinear fashion with $N$, as the effective collective free space decay rate $\gamma\ts{eff}$ can be dramatically suppressed. We propose an example for the implementation of phase imprinting using higher order transverse cavity modes leading to the preferential excitation of subradiant collective states. The effect is directly observable by homodyne detection of the cavity output, displayed both in amplitude and phase antiresonant behavior~\cite{sames2014antiresonance,rice1996cavity}. As opposed to the strong coupling regime exploited in~\cite{sames2014antiresonance}, this paper considers the bad cavity regime $\kappa\gg g_1\sqrt{N}$ where one typically expects modest antiresonance phase shifts. Because of the collectively increased effective cooperativity, very narrow antiresonances occur accompanied by extremely fast and large phase shift switches rendering such a system perfect for high resolution spectroscopy.

\section{Model}
Let us consider an ordered ensemble of quantum emitters modeled as two-level systems with ground state $\ket{g}_i$ and an excited state $\ket{e}_i$ (split by frequency $\omega\ts{e}$) located at $\textbf{r}_i$ (for $i=1,...,N$) (see Fig.~\ref{fig1}). The levels are connected by individual Pauli raising and lowering operators $ \sigma_i^\pm$ with $\sigma_i^x=\sigma_i^+ + \sigma_i^-$, $\sigma_i^y=-i(\sigma_i^+-\sigma_i^-) $, and $\sigma_i^z= \sigma_i^+ \sigma_i^- - \sigma_i^-\sigma_i^+$. The emitters are embedded in a static 2D support, transversely placed in the center plane of a single higher order transverse electromagnetic (TEM) mode at frequency $\omega_\text{c}$  (see~\fref{fig1}). At position $(\textbf{r},z)$ along the cavity axis $z$, the electric field operator is proportional to $a\cos(kz) f(\textbf{r}) \boldsymbol{\epsilon}_y$, where $a$ is the annihilation operator of the cavity mode, $k=\omega\ts{c}/c$, $f(\textbf{r})$ is the transverse spatial mode profile, and $\boldsymbol{\epsilon}_y$ denotes linear polarization in the $y$ direction. The cavity is laser driven at frequency $\omega\ts{l}$ with power $P$ through one mirror. In a frame rotating at $\omega\ts{l}$, the dynamics of the mode of interest is described by
\begin{align}
H_\text{c} = \hbar\Delta\ts{c}a^\dag a + i\hbar\eta\left(a^\dag - a\right),
\end{align}
where $\Delta\ts{c}=\omega\ts{c} - \omega\ts{l}$ and $\eta=\sqrt{2P\kappa/(\hbar\omega\ts{l})}$. Cavity damping with decay rate $\kappa$ occurs via the collapse operator $a$.

At dense spacing ($|\textbf{r}_i - \textbf{r}_{i+1}| < \lambda\ts{e}$), one has to account for the direct emitter-emitter interactions via the transition dipole moments $\boldsymbol{\mu}_i$. The collective dynamics is governed by the free Hamiltonian $H\ts{e}$ and the collective part $H\ts{dip}$,
\begin{align}
H\ts{e} + H\ts{dip} = \hbar\Delta_\text{e}\sum_i\sigma_i^+\sigma_i^- + \hbar\sum_{i, j : i\neq j}\Omega_{ij}\sigma_i^+\sigma_j^-,
\end{align}
where $\Delta\ts{e} = \omega\ts{e} - \omega\ts{l}$, and $\Omega_{ij}$ is the strength of the coherent dipole-dipole interaction between emitters $i$ and $j$ (see Appendix). Moreover, the incoherent collective dynamics leads to mutual decay rates $\gamma_{ij}$ that can be accounted for with the Lindblad superoperator~\cite{lehmberg1970radiation}
\begin{align}
\mathcal{L}\ts{e} [\rho] = \sum_{i,j}\gamma_{ij}\left(2\sigma_i^-\rho\sigma_j^+ - \sigma_i^+\sigma_j^-\rho - \rho\sigma_i^+\sigma_j^-\right).
\end{align}
\indent In the single cavity mode limit, the interaction is described by the Tavis-Cummings Hamiltonian
\begin{align}
H\ts{int} = \hbar\sum_i g_i\left(a^\dag\sigma_i^- + a\sigma_i^+\right),
\end{align}
where the coupling strength $g_i$ of an emitter at position $(\textbf{r}_i, z_i)$ is proportional to $\cos(kz_i) f(x_i,y_i) \boldsymbol{\epsilon}_y\cdot \boldsymbol{\mu}_i$.

The complete dynamics of the system with density matrix $\rho$ are then described by the master equation
\begin{align}
\dot{\rho} &= \frac{i}{\hbar}[\rho,H] + \mathcal{L}\ts{c}[\rho] + \mathcal{L}\ts{e}[\rho],
\end{align}
where $H = H\ts{c} + H\ts{e} + H\ts{dip} + H\ts{int}$ and $\mathcal{L}\ts{c}[\rho] = \kappa\left(2a\rho a^\dag - a^\dag a \rho - \rho a^\dag a\right)$.
Equivalently, the dynamics can be described via quantum Langevin equations (QLE)~\cite{gardiner2004quantum} (see Appendix).

\section{Single emitter antiresonance}
We consider a reference system with a single emitter in the low excitation limit $\langle\sigma^z_i\rangle\approx -1$ where a linear coupled set of QLEs can be derived. For a resonant interaction (i.e. $\Delta\equiv\Delta\ts{c}=\Delta\ts{e}$), this leads to the following mean field equations:
\begin{align}
\braket{\dot{a}} &= -(\kappa +i\Delta)\braket{a} + \eta -ig\braket{\sigma^-},
\label{single_HE1}
\\
\braket{\dot{\sigma}^-} &= -(\gamma+i\Delta)\braket{\sigma^-} - i g \braket{a}.
\label{single_HE2}
\end{align}
These equations exhibit the phenomenon of atomic antiresonances \cite{alsing1992suppression,sames2014antiresonance}, where the resonantly driven atomic dipole oscillates in a way to counteract the cavity drive and leads to a minimum of transmission~\cite{zippilli2004suppression}. We analyze its dependence on $\gamma$ by studying the steady-state amplitude transmission $t$, which is proportional to the output field amplitude $t=\kappa\braket{a}/\eta$. It reads
\begin{align}
t &= \frac{\kappa}{i\Delta+\kappa+g^2/(i\Delta + \gamma)}.
\end{align}
The transmitted intensity is $T=|t|^2$ and the relative phase shift caused by the emitter is $\phi-\phi\ts{c}$, where
$\phi=\text{Arg}(t) = \arctan\left(\Im\{t\}/\Re\{t\}\right)$
and $\phi\ts{c}=-\arctan\left(\Delta\ts{c}/\kappa\right)$ is the phase shift of the bare cavity. The detection of the relative phase shift can be done by homodyne detection and analysis of the output field quadratures. Scanning the laser frequency ($\Delta$), we find that the coherent transmitted intensity through the cavity contains an antiresonance dip around $\Delta=0$ with a corresponding jump in the phase shift (see \fref{fig1}). Fitting the antiresonance with a Lorentzian (see Appendix), we find a depth of $1-T(\Delta=0)=C_1(C_1+2)/(C_1+1)^2$, and a width that can be approximated by $\gamma(C_1 + 1)=g_1^2/\kappa + \gamma$ (for a regime where both $g_1,\gamma\ll\kappa$). An almost vanishing transmission is, then, a signature of reaching a regime of strong cooperativity ($C_1\gg 1$).

\section{Collective antiresonance of emitter arrays}
As $C_1$ is independent of $\mu$, an emitter with a larger dipole moment will only broaden the antiresonance. For coupled emitter arrays, this is, however, no longer valid, and one can design the radiative properties of the ensemble. For \textit{collective} subradiant resonances of an array the free space emission is suppressed, while we still can optimize the coupling to the cavity mode. This generates extremely sharp and deep antiresonances accompanied by a fast and large phase change within a narrow frequency range. The immediate upshot of this regime is a dramatically enhanced effective cooperativity, which renders it an ideal configuration for high resolution spectroscopy.

The set of coupled QLEs for many emitters can be cast in vector form
\begin{align}
\dot{\braket{a}} &= -i\Delta_\text{c}\braket{a}+\eta - i\textbf{G}^\intercal\braket{\boldsymbol{\sigma}} - \kappa \braket{a},
\label{HE1}
\\
\dot{\braket{\boldsymbol{\sigma}}} &= -i \Delta\ts{e} \braket{\boldsymbol{\sigma}} - i \boldsymbol{\Omega}\braket{\boldsymbol{\sigma}} - i \textbf{G}\braket{a} - \boldsymbol{\Gamma}\braket{\boldsymbol{\sigma}},
\label{HE2}
\end{align}
where, now, $\boldsymbol{\sigma}$ and $\textbf{G}$ are column vectors with entries $\sigma_i^-$ and $g_i$. The matrices $\boldsymbol{\Omega}$ and $\boldsymbol{\Gamma}$ have the elements $\Omega_{ij}$ and $\gamma_{ij}$. In steady state the transmission coefficient for the cavity amplitude reads
\begin{align}
t  &= \frac{\kappa}{i\Delta\ts{c}+\kappa + \textbf{G}^\intercal\textbf{G}/[i\Delta_\text{eff}(\Delta\ts{e}) + \gamma_\text{eff}(\Delta\ts{e})]},
\end{align}
where the effective $\Delta\ts{e}$-dependent collective energy shifts and linewidths are derived from the matrix
\begin{align}
\textbf{M}(\Delta\ts{e})= i\Delta\ts{e}\mathbbm{1}+i\boldsymbol{\Omega}+\boldsymbol{\Gamma}
\end{align}
as real and imaginary parts
\begin{align}
\label{delta_eff}
\Delta_\text{eff}(\Delta\ts{e}) &= \Im\left\{\frac{\textbf{G}^\intercal\textbf{G}}{\textbf{G}^\intercal\textbf{M}^{-1}(\Delta\ts{e})\textbf{G}}\right\},
\\
\gamma_\text{eff}(\Delta\ts{e}) &= \Re\left\{\frac{\textbf{G}^\intercal\textbf{G}}{\textbf{G}^\intercal \textbf{M}^{-1}(\Delta\ts{e})\textbf{G}}\right\}.
\label{gamma_eff}
\end{align}
In analogy to the single emitter case, we can define an effective $N$-emitter cooperativity by
\begin{align}
\label{CN_def}
C\ts{eff}(\Delta\ts{e}) = \frac{\textbf{G}^\intercal\textbf{G}}{\kappa\gamma_\text{eff}(\Delta\ts{e})}.
\end{align}
This equation provides a main message of the paper, as it shows that the numerator and denominator no longer share the same dependency on $\mu$. As $\gamma_\text{eff}$ is not a natural constant of the ensemble, but strongly dependent on the relative positioning and phase of individual emitters, one can reach subradiant states with $\gamma_\text{eff}\ll\gamma$. By proper design of the cavity transverse field amplitude profile, the numerator can, at the same time, be maximized, resulting in a scaling up of $C\ts{eff}$ well above the independent emitter case $N g_1^2/(\kappa \gamma)$.

\section{Two emitters}
Let us elucidate the mechanism in the two emitter case, with adjustable separation $d=|\textbf{r}_1-\textbf{r}_2|$. We distinguish two fundamentally different cases: (i) uniform coupling $\textbf{G}=(g,~g)^\intercal$ and (ii) opposite coupling $\textbf{G}=(g,-g)^\intercal$, resulting in $\textbf{G}^\intercal\textbf{G}=2g^2$ for both cases. The matrix of interactions can be diagonalized with eigenvalues $i(\Delta\ts{e}\pm\Omega_{12})+(\gamma\pm\gamma_{12})$, signaling the presence of collective super- and subradiant states ($\gamma \pm \gamma_{12}$) shifted by $\pm \Omega_{12}$ from the emitter resonance $\omega\ts{e}$ (the positive sign corresponds to uniform coupling).

\begin{figure}[t]
\includegraphics[width=\columnwidth]{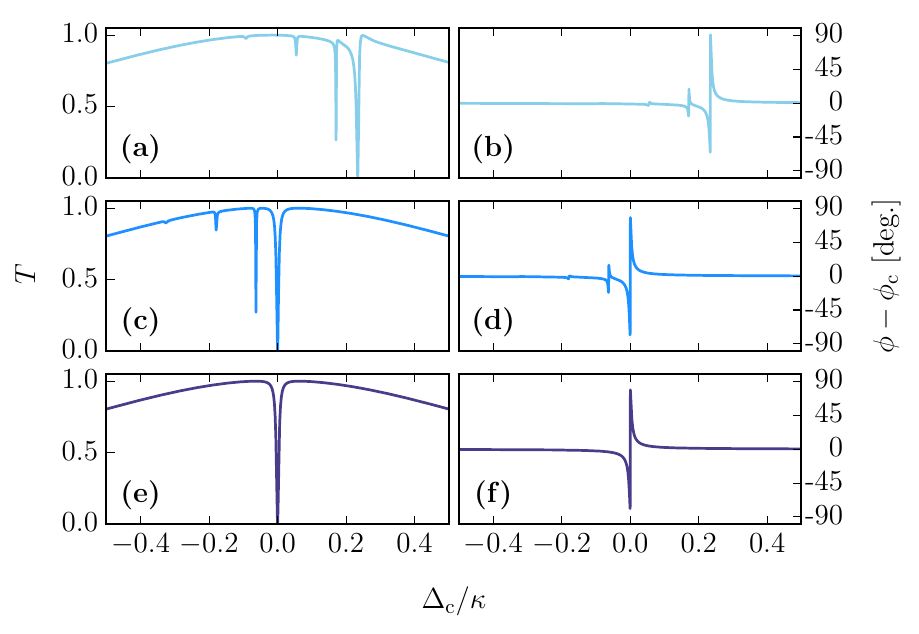}
\caption{\emph{Targeting antiresonances}. (a), (c), and (e) show the cavity intensity transmission and (b), (d), and (f) the corresponding phase for a scan of the laser frequency. The upper row corresponds to $\omega\ts{c}=\omega\ts{e}$ and asymmetric coupling $g_i=(-1)^i g$; the middle row illustrates frequency selection as $\omega\ts{c}\simeq\omega\ts{m=N}$. Finally, we also match the symmetry to the state corresponding to the subradiant antiresonance (see Appendix). The parameters are $N=10$, $g=\kappa/50$, $\gamma=\kappa/40$ with a chain separation of $d=0.08 \lambda\ts{e}$.}
\label{fig2}
\end{figure}

In the extreme case, where $d \ll \lambda_e$, the mutual decay approaches $\gamma_{12}\to\gamma$, and the effective cooperativity reaches (i) $C\ts{eff}\to g^2/(\kappa\gamma)$ and (ii) $C\ts{eff}\to\infty$, respectively. To account for dipole-dipole energy shifts, we impose $\Delta\ts{eff}(\delta)=0$ and, subsequently, tune the cavity such that $\omega\ts{c}=\omega\ts{e} - \delta$; i.e. we match the cavity to the shifted collective emitter resonance. For two emitters, the imposed resonance condition yields $\delta = \pm \Omega_{12}$. The resulting depth and width of each antiresonance is $C\ts{eff}(C\ts{eff}+2)/(C\ts{eff}+1)^2$ and $\gamma\ts{eff}(C\ts{eff} + 1)$, respectively. Hence, for $d\to 0$, we have (i) an antiresonance depth as for the single emitter but twice the width (superradiance), and (ii) an antiresonance that has a depth of 1 and a width of $2g^2/\kappa$ (subradiance). While the width of the antiresonance is still limited by $g$, the phase switch bandwidth is independent of $g$. This is a direct measure of the subradiance as the slope of the phase switch in this limit is $1/\gamma\ts{eff}$ (see Appendix). The result is reminiscent of the one in~\cite{sames2014antiresonance}, however, in a very different and less stringent regime, where only weak coupling is required and where usually moderate phase shifts are expected; in contrast, for $\gamma\ts{eff}\to 0$, the phase even exhibits a $\pi$ phase change within an extremely narrow frequency range, since in this regime, $\lim_{\Delta\to 0^\pm}\left(\phi - \phi\ts{c}\right) = \pm\pi/2$.

\section{Addressing collective subradiant states}
The above results can be generalized to $N$ emitters in an equidistant chain configuration ($d=|r_{i+1}-r_i|$). Analytical considerations can be made under a nearest neighbor approximation for $H\ts{dip}$ in the single-excitation regime, very well justified at small interemitter distances and weak driving. Diagonalization of $H\ts{dip}$ gives rise to an $N$-band problem with energies $\omega\ts{m} = \omega\ts{e}+2\Omega_{12}\cos[m\pi/(N+1)]$ for $m$ running from $1$ to $N$. The Lindblad term then shows a ranking of levels from superradiant ($m=1$) to very subradiant ($m=N$) for $d\ll\lambda\ts{e}/2$~\cite{plankensteiner2015selective}. Moreover, the eigenvectors $|m\rangle=\sqrt{2/(N+1)}\sum_j\sin[mj\pi/(N+1)]\sigma_j^+\ket{g}^{\otimes N}$ have a specific geometry with almost full symmetry ($m=1$) to almost full asymmetry ($m=N$). The two distinct cases involving uniform $\textbf{G}=(g,~g,...)^\intercal$ and opposite couplings $\textbf{G}=(g,-g,...)^\intercal$ then almost perfectly address these fully symmetric $|m=1\rangle$ and asymmetric $|m=N\rangle$ states.

Illustrated in Figs.~\ref{fig2}(a) and \ref{fig2}(b) is a scan of the collective resonances of a ten-emitter chain with $g_i=(-1)^i g$ and $\omega\ts{c}=\omega\ts{e}$. Both the dip and phase show an off-resonant selection of collective subradiant states. We then selectively target a given state by fitting the cavity resonance to its energy as shown in Figs.~\ref{fig2}(c) and \ref{fig2}(d). To achieve this, we focus around the state $|m=N\rangle$ with energy $\omega\ts{m=N}$ and we recalculate the state's energy by imposing $\Delta\ts {eff} (\delta)=0$ after which we set $\omega\ts{c}=\omega\ts{e}-\delta$. We note that, as opposed to the two-emitter case, we cannot find an analytical value for $\delta$ but solve for it numerically. It corresponds to a value close to $\omega\ts{m=N}-\omega\ts{e}$. Finally, we compare the results to an ideal procedure where the components of $\textbf{G}$ are chosen such that they match the geometry of the target state [Figs.~\ref{fig2}(e) and \ref{fig2}(f)].

\begin{figure}[t]
\includegraphics[width=\columnwidth]{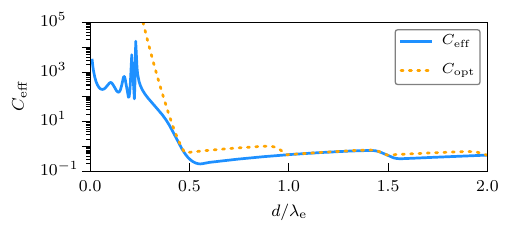}
\caption{\emph{Scaling of effective cooperativity}. $C\ts{eff}$ for $N=10$ as a function of $d/\lambda\ts{e}$. The effective cooperativity (blue, solid curve) is compared to an idealized case of perfect subradiance (yellow, dashed curve). We used $\Delta\ts{c}=\Delta\ts{eff}=0$, $g=\kappa/30$, and $\gamma=\kappa/40$.}
\label{fig3}
\end{figure}

The characteristics of the antiresonances can be quantified by $C\ts{eff}$ (see~\fref{fig3}). As above, we assume the asymmetric cavity field profile with $\textbf{G}=(g,-g,...)^\intercal$ and make a reference plot $C\ts{opt}$ as a function of $d$. The optimal cooperativity $C\ts{opt}$ is obtained from Eq.~\eqref{CN_def} by substituting the decay rate with the minimal eigenvalue of $\boldsymbol{\Gamma}$~\cite{ostermann2014protected}. In reality, owing to imperfect phase matching to the most subradiant state as well as to the inherent level shifts brought on by the dipole-dipole interactions, the effective gain is more modest. Nevertheless, as suggested by the blue (solid) curve in~\fref{fig3}, for $d<0.5\lambda\ts{e}$ the enhancement is considerably larger than in the noninteracting quantum emitters case.

\section{Subradiance using transverse phase gradients}
\begin{figure}[b]
\includegraphics[width=\columnwidth]{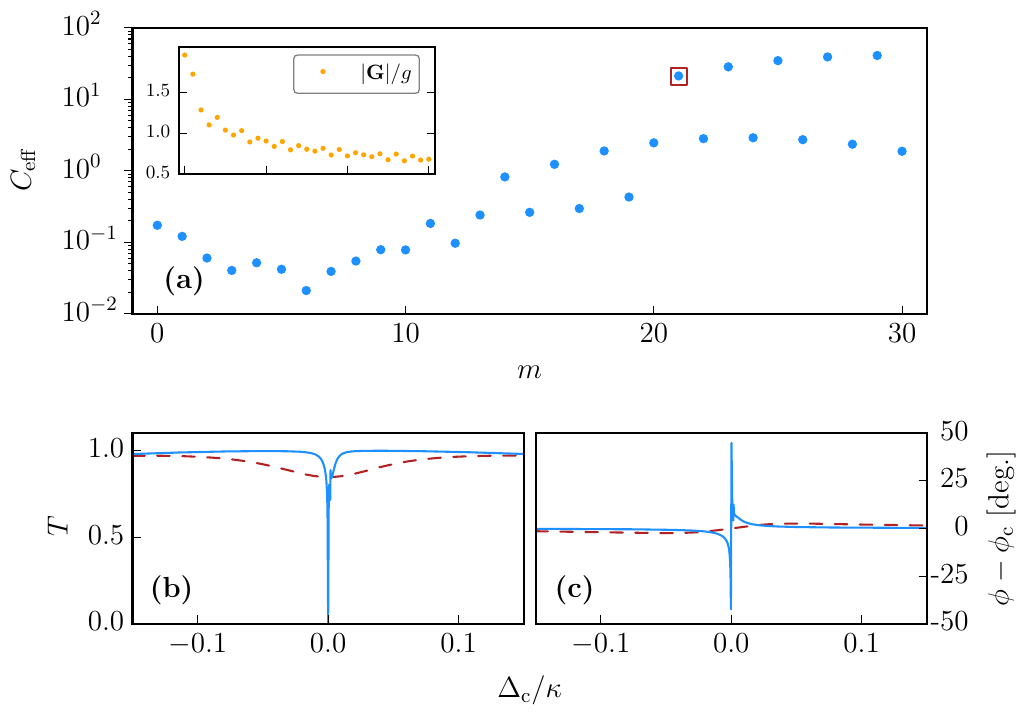}
\caption{\emph{Targeting subradiance via transverse mode driving}. (a) Effective cooperativity as a function of $m$ for $N=10$ and under illumination with TEM$\ts{m0}$ mode. The inset shows a decrease in cavity coupling $|\mathbf{G}|$. (b),(c) Comparison of antiresonance signatures for TEM$\ts{00}$ addressing (red, dashed line) vs TEM$_{m0}$ addressing (blue, solid line) with $m=21$, corresponding to the point enclosed in the red box in (a). The parameters are $d=0.2\lambda\ts{e}$, $w=\lambda\ts{e}$, $g=\kappa/30$, $\gamma=\kappa/40$, and $\Delta\ts{eff}=0$ at $\Delta\ts{c}=0$.}
\label{fig4}
\end{figure}
While, in practice, individual phase imprinting on the subwavelength scale is not a trivial task, we present a theoretical illustration using 1D or 2D ensembles transversely placed in the center of a cavity, in the focal point of a higher order TEM mode. In the plane of the emitters, the field profile of a Gaussian-Hermite mode of order $m,n$ is $f\left(x,y\right)= A_{mn} H_m\left(\sqrt{2}x/w\right)H_n\left(\sqrt{2}y/w\right)e^{-(x^2 + y^2)/w^2}$. Here, $H_n(x)$ is the $n$th Hermite polynomial, $w$ is the waist of the beam in the center of the cavity assuming a perfectly symmetric cavity and $A_{mn}=\sqrt{2/(\pi 2^{(m+n)}m!n!)}$. Higher order TEM modes exhibit multiple extrema of opposite signs in the transverse profile. For a sufficiently small $w$ (of the order of $\lambda\ts{e}$), adjacent extrema can be closely spaced (for a TEM$\ts{m0}$ mode around $w/\sqrt{m}$) resulting in the desired coupling asymmetry. Note that, in reality, owing to the diffraction limit, optical cavities might not be stable under high transverse mode operation, in which case, an alternative stability regime has to be found.\\
\indent We illustrate the phase imprinting mechanism for a chain illuminated by a TEM$\ts{m0}$ mode with increasing $m$ in~\fref{fig4}(a), where now $g$ is the coupling strength in the center of the $\text{TEM}_{00}$ mode. While, for small $m$, the effective cooperativity decreases (owing to a decrease in $|\mathbf{G}|$), at higher $m$ the alternating field phases are partially addressing asymmetric collective states of high robustness resulting in a considerably enhanced effective cooperativity. The very sharp cavity response for a fixed mode $m=21$ is shown in Figs.~\ref{fig4}(b) and \ref{fig4}(c) in comparison to the modest results expected for a TEM$_{00}$ illumination. Moreover, we have numerically investigated 2D geometries as well and found, for example, in the case of a $3\times3$ square array with $d= 0.2\lambda\ts{e}=w$ and $g=\kappa/20=2\gamma$ (as depicted in~\fref{fig1}), an enhancement of effective cooperativity from the bare value $Ng^2/(\kappa\gamma) = 0.9$ to $C\ts{eff}\approx 80.1$.

\section{Conclusions}
Tailoring the \textit{collective} dissipative dynamics of $N$ dipole coupled emitters can lead to high effective cooperativity even in the regime $Ng^2/(\kappa\gamma)\ll1$. The immediate consequence is the occurrence of a narrow antiresonance dip with fast spectral phase switching without the need of strong individual coupling as in Ref.~\cite{sames2014antiresonance}. As it applies to narrow atomic transitions, it hints towards applications for precision spectroscopy and quantum network characterization. As opposed to using a lossy cavity field as an engineered bath leading to superradiance as in Ref.~\cite{mascarenhas2013cooperativity}, we only considered the naturally occurring environment provided by the free space radiation modes. The regime treated here is perturbative; i.e. the emitters do not modify the bare mode functions of the cavity mode. Increasing the collective scattering rate close to unity~\cite{jenkins2012controlled,jenkins2013metamaterial,bettles2015cooperative, bettles2016cooperative, bettles2016enhanced, shahmoon2016cooperativity} should result in an interesting regime of cavity QED where the cavity mode functions are strongly modified by a relatively modest number of emitters. A dynamical regime can occur and be exploited for hybrid optomechanical applications~\cite{juan2016near,dantan2014hybrid} with emitters implanted on vibrating membranes. Stronger phonon-photon interactions could be designed that benefit from narrow collective resonances~\cite{dantan2014hybrid}. Similar considerations can be used to analyze metamaterial arrays, where classical analogues of subradiant states are also experimentally seen~\cite{jenkins2016many}.

\begin{acknowledgments}
We acknowledge financial support by the Austrian Science Fund (FWF) within the Innsbruck Doctoral School DK-ALM with Grant No. W1259-N27 (D.~P.) and Project No. P29318-N27 (H.~R) as well as from the Aarhus Universitets Forskningsfond and the Max Planck Society (C.~G). We also acknowledge useful discussions with Robert Bettles. The graphs were done with the open source library MATPLOTLIB~\cite{hunter2007matplotlib}.
\end{acknowledgments}

\bibliographystyle{apsrev4-1-custom}
\bibliography{AntiResRef}

\begin{thebibliography}{31}%
\makeatletter
\providecommand \@ifxundefined [1]{%
 \@ifx{#1\undefined}
}%
\providecommand \@ifnum [1]{%
 \ifnum #1\expandafter \@firstoftwo
 \else \expandafter \@secondoftwo
 \fi
}%
\providecommand \@ifx [1]{%
 \ifx #1\expandafter \@firstoftwo
 \else \expandafter \@secondoftwo
 \fi
}%
\providecommand \natexlab [1]{#1}%
\providecommand \enquote  [1]{``#1''}%
\providecommand \bibnamefont  [1]{#1}%
\providecommand \bibfnamefont [1]{#1}%
\providecommand \citenamefont [1]{#1}%
\providecommand \href@noop [0]{\@secondoftwo}%
\providecommand \href [0]{\begingroup \@sanitize@url \@href}%
\providecommand \@href[1]{\@@startlink{#1}\@@href}%
\providecommand \@@href[1]{\endgroup#1\@@endlink}%
\providecommand \@sanitize@url [0]{\catcode `\\12\catcode `\$12\catcode
  `\&12\catcode `\#12\catcode `\^12\catcode `\_12\catcode `\%12\relax}%
\providecommand \@@startlink[1]{}%
\providecommand \@@endlink[0]{}%
\providecommand \url  [0]{\begingroup\@sanitize@url \@url }%
\providecommand \@url [1]{\endgroup\@href {#1}{\urlprefix }}%
\providecommand \urlprefix  [0]{URL }%
\providecommand \Eprint [0]{\href }%
\providecommand \doibase [0]{http://dx.doi.org/}%
\providecommand \selectlanguage [0]{\@gobble}%
\providecommand \bibinfo  [0]{\@secondoftwo}%
\providecommand \bibfield  [0]{\@secondoftwo}%
\providecommand \translation [1]{[#1]}%
\providecommand \BibitemOpen [0]{}%
\providecommand \bibitemStop [0]{}%
\providecommand \bibitemNoStop [0]{.\EOS\space}%
\providecommand \EOS [0]{\spacefactor3000\relax}%
\providecommand \BibitemShut  [1]{\csname bibitem#1\endcsname}%
\let\auto@bib@innerbib\@empty
\bibitem [{\citenamefont {Haroche} and \citenamefont
  {Kleppner}(1989)}]{haroche1989cavity}%
  \BibitemOpen
  \bibfield  {author} {\bibinfo {author} {\bibfnamefont {S.}~\bibnamefont
  {Haroche}} and \bibinfo {author} {\bibfnamefont {D.}~\bibnamefont
  {Kleppner}}, }\bibfield  {title} {\enquote {\bibinfo {title} {Cavity
  {Q}uantum {E}lectrodynamics},} }\href@noop {} {\bibfield  {journal} {\bibinfo
   {journal} {Phys. Today} }\textbf {\bibinfo {volume} {42}}, \bibinfo {pages}
  {24--30} (\bibinfo {year} {1989})}\BibitemShut {NoStop}%
\bibitem [{\citenamefont {Walther} \emph {et~al.}(2006)\citenamefont {Walther},
  \citenamefont {Varcoe}, \citenamefont {Englert}, and \citenamefont
  {Becker}}]{walther2006cavity}%
  \BibitemOpen
  \bibfield  {author} {\bibinfo {author} {\bibfnamefont {H.}~\bibnamefont
  {Walther}}, \bibinfo {author} {\bibfnamefont {B.~T.} \bibnamefont {Varcoe}},
  \bibinfo {author} {\bibfnamefont {B.}~\bibnamefont {Englert}},  and \bibinfo
  {author} {\bibfnamefont {T.}~\bibnamefont {Becker}}, }\bibfield  {title}
  {\enquote {\bibinfo {title} {Cavity {Q}uantum {E}lectrodynamics},} }\href
  {\doibase 10.1088/0034-4885/69/5/R02} {\bibfield  {journal} {\bibinfo
  {journal} {Rep. Prog. Phys.} }\textbf {\bibinfo {volume} {69}}, \bibinfo
  {pages} {1325} (\bibinfo {year} {2006})}\BibitemShut {NoStop}%
\bibitem [{\citenamefont {Kimble}(1998)}]{kimble1998strong}%
  \BibitemOpen
  \bibfield  {author} {\bibinfo {author} {\bibfnamefont {H.~J.} \bibnamefont
  {Kimble}}, }\bibfield  {title} {\enquote {\bibinfo {title} {Strong
  interactions of single atoms and photons in cavity {QED}},} }\href@noop {}
  {\bibfield  {journal} {\bibinfo  {journal} {Physica Scripta} }\textbf
  {\bibinfo {volume} {1998}}, \bibinfo {pages} {127} (\bibinfo {year}
  {1998})}\BibitemShut {NoStop}%
\bibitem [{\citenamefont {Birnbaum} \emph {et~al.}(2005)}]{birnbaum2005photon}%
  \BibitemOpen
  \bibfield  {author} {\bibinfo {author} {\bibfnamefont {K.~M.} \bibnamefont
  {Birnbaum}} \emph {et~al.}, }\bibfield  {title} {\enquote {\bibinfo {title}
  {Photon blockade in an optical cavity with one trapped atom},} }\href
  {\doibase 10.1038/nature03804} {\bibfield  {journal} {\bibinfo  {journal}
  {Nature} }\textbf {\bibinfo {volume} {436}}, \bibinfo {pages} {87--90}
  (\bibinfo {year} {2005})}\BibitemShut {NoStop}%
\bibitem [{\citenamefont {Yoshie} \emph {et~al.}(2004)}]{yoshie2004vacuum}%
  \BibitemOpen
  \bibfield  {author} {\bibinfo {author} {\bibfnamefont {T.}~\bibnamefont
  {Yoshie}} \emph {et~al.}, }\bibfield  {title} {\enquote {\bibinfo {title}
  {Vacuum {R}abi splitting with a single quantum dot in a photonic crystal
  nanocavity},} }\href {\doibase 10.1038/nature03119} {\bibfield  {journal}
  {\bibinfo  {journal} {Nature} }\textbf {\bibinfo {volume} {432}}, \bibinfo
  {pages} {200--203} (\bibinfo {year} {2004})}\BibitemShut {NoStop}%
\bibitem [{\citenamefont {Casanova} \emph {et~al.}(2010)\citenamefont
  {Casanova}, \citenamefont {Romero}, \citenamefont {Lizuain}, \citenamefont
  {Garc\'{\i}a-Ripoll}, and \citenamefont {Solano}}]{casanova2010deep}%
  \BibitemOpen
  \bibfield  {author} {\bibinfo {author} {\bibfnamefont {J.}~\bibnamefont
  {Casanova}}, \bibinfo {author} {\bibfnamefont {G.}~\bibnamefont {Romero}},
  \bibinfo {author} {\bibfnamefont {I.}~\bibnamefont {Lizuain}}, \bibinfo
  {author} {\bibfnamefont {J.~J.} \bibnamefont {Garc\'{\i}a-Ripoll}},  and
  \bibinfo {author} {\bibfnamefont {E.}~\bibnamefont {Solano}}, }\bibfield
  {title} {\enquote {\bibinfo {title} {Deep strong coupling regime of the
  jaynes-cummings model},} }\href {\doibase 10.1103/PhysRevLett.105.263603}
  {\bibfield  {journal} {\bibinfo  {journal} {Phys. Rev. Lett.} }\textbf
  {\bibinfo {volume} {105}}, \bibinfo {pages} {263603} (\bibinfo {year}
  {2010})}\BibitemShut {NoStop}%
\bibitem [{\citenamefont {Vuleti{\'c}} \emph {et~al.}(2001)\citenamefont
  {Vuleti{\'c}}, \citenamefont {Chan}, and \citenamefont
  {Black}}]{vuletic2001three}%
  \BibitemOpen
  \bibfield  {author} {\bibinfo {author} {\bibfnamefont {V.}~\bibnamefont
  {Vuleti{\'c}}}, \bibinfo {author} {\bibfnamefont {H.~W.} \bibnamefont
  {Chan}},  and \bibinfo {author} {\bibfnamefont {A.~T.} \bibnamefont {Black}},
  }\bibfield  {title} {\enquote {\bibinfo {title} {Three-dimensional cavity
  {D}oppler cooling and cavity sideband cooling by coherent scattering},}
  }\href@noop {} {\bibfield  {journal} {\bibinfo  {journal} {Phys. Rev. A}
  }\textbf {\bibinfo {volume} {64}}, \bibinfo {pages} {033405} (\bibinfo {year}
  {2001})}\BibitemShut {NoStop}%
\bibitem [{\citenamefont {Plankensteiner} \emph {et~al.}(2015)\citenamefont
  {Plankensteiner}, \citenamefont {Ostermann}, \citenamefont {Ritsch}, and
  \citenamefont {Genes}}]{plankensteiner2015selective}%
  \BibitemOpen
  \bibfield  {author} {\bibinfo {author} {\bibfnamefont {D.}~\bibnamefont
  {Plankensteiner}}, \bibinfo {author} {\bibfnamefont {L.}~\bibnamefont
  {Ostermann}}, \bibinfo {author} {\bibfnamefont {H.}~\bibnamefont {Ritsch}},
  and \bibinfo {author} {\bibfnamefont {C.}~\bibnamefont {Genes}}, }\bibfield
  {title} {\enquote {\bibinfo {title} {Selective protected state preparation of
  coupled dissipative quantum emitters},} }\href {\doibase 10.1038/srep16231}
  {\bibfield  {journal} {\bibinfo  {journal} {Sci. Rep.} }\textbf {\bibinfo
  {volume} {5}}, \bibinfo {pages} {16231} (\bibinfo {year} {2015})}\BibitemShut
  {NoStop}%
\bibitem [{\citenamefont {Bettles} \emph {et~al.}(2015)\citenamefont {Bettles},
  \citenamefont {Gardiner}, and \citenamefont
  {Adams}}]{bettles2015cooperative}%
  \BibitemOpen
  \bibfield  {author} {\bibinfo {author} {\bibfnamefont {R.~J.} \bibnamefont
  {Bettles}}, \bibinfo {author} {\bibfnamefont {S.~A.} \bibnamefont
  {Gardiner}},  and \bibinfo {author} {\bibfnamefont {C.~S.} \bibnamefont
  {Adams}}, }\bibfield  {title} {\enquote {\bibinfo {title} {Cooperative
  ordering in lattices of interacting two-level dipoles},} }\href {\doibase
  10.1103/PhysRevA.92.063822} {\bibfield  {journal} {\bibinfo  {journal} {Phys.
  Rev. A} }\textbf {\bibinfo {volume} {92}}, \bibinfo {pages} {063822}
  (\bibinfo {year} {2015})}\BibitemShut {NoStop}%
\bibitem [{\citenamefont {Bettles} \emph
  {et~al.}(2016{\natexlab{a}})\citenamefont {Bettles}, \citenamefont
  {Gardiner}, and \citenamefont {Adams}}]{bettles2016cooperative}%
  \BibitemOpen
  \bibfield  {author} {\bibinfo {author} {\bibfnamefont {R.~J.} \bibnamefont
  {Bettles}}, \bibinfo {author} {\bibfnamefont {S.~A.} \bibnamefont
  {Gardiner}},  and \bibinfo {author} {\bibfnamefont {C.~S.} \bibnamefont
  {Adams}}, }\bibfield  {title} {\enquote {\bibinfo {title} {Cooperative
  eigenmodes and scattering in one-dimensional atomic arrays},} }\href
  {\doibase 10.1103/PhysRevA.94.043844} {\bibfield  {journal} {\bibinfo
  {journal} {Phys. Rev. A} }\textbf {\bibinfo {volume} {94}}, \bibinfo {pages}
  {043844} (\bibinfo {year} {2016}{\natexlab{a}})}\BibitemShut {NoStop}%
\bibitem [{\citenamefont {Bettles} \emph
  {et~al.}(2016{\natexlab{b}})\citenamefont {Bettles}, \citenamefont
  {Gardiner}, and \citenamefont {Adams}}]{bettles2016enhanced}%
  \BibitemOpen
  \bibfield  {author} {\bibinfo {author} {\bibfnamefont {R.~J.} \bibnamefont
  {Bettles}}, \bibinfo {author} {\bibfnamefont {S.~A.} \bibnamefont
  {Gardiner}},  and \bibinfo {author} {\bibfnamefont {C.~S.} \bibnamefont
  {Adams}}, }\bibfield  {title} {\enquote {\bibinfo {title} {Enhanced optical
  cross section via collective coupling of atomic dipoles in a 2{D} array},}
  }\href {\doibase 10.1103/PhysRevLett.116.103602} {\bibfield  {journal}
  {\bibinfo  {journal} {Phys. Rev. Lett.} }\textbf {\bibinfo {volume} {116}},
  \bibinfo {pages} {103602} (\bibinfo {year} {2016}{\natexlab{b}})}\BibitemShut
  {NoStop}%
\bibitem [{\citenamefont {Shahmoon} \emph {et~al.}(2016)\citenamefont
  {Shahmoon}, \citenamefont {Wild}, \citenamefont {Lukin}, and \citenamefont
  {Yelin}}]{shahmoon2016cooperativity}%
  \BibitemOpen
  \bibfield  {author} {\bibinfo {author} {\bibfnamefont {E.}~\bibnamefont
  {Shahmoon}}, \bibinfo {author} {\bibfnamefont {D.~S.} \bibnamefont {Wild}},
  \bibinfo {author} {\bibfnamefont {M.~D.} \bibnamefont {Lukin}},  and \bibinfo
  {author} {\bibfnamefont {S.~F.} \bibnamefont {Yelin}}, }\bibfield  {title}
  {\enquote {\bibinfo {title} {Cooperative resonances in light scattering from
  two-dimensional atomic arrays},} }\href {https://arxiv.org/abs/1610.00138}
  {\bibfield  {journal} {\bibinfo  {journal} {arxiv:1610.00138} } (\bibinfo
  {year} {2016})}\BibitemShut {NoStop}%
\bibitem [{\citenamefont {Perczel} \emph {et~al.}(2017)\citenamefont {Perczel},
  \citenamefont {Borregaard}, \citenamefont {Chang}, \citenamefont {Pichler},
  \citenamefont {Yelin}, \citenamefont {Zoller}, and \citenamefont
  {Lukin}}]{perczel2017topological}%
  \BibitemOpen
  \bibfield  {author} {\bibinfo {author} {\bibfnamefont {J.}~\bibnamefont
  {Perczel}}, \bibinfo {author} {\bibfnamefont {J.}~\bibnamefont {Borregaard}},
  \bibinfo {author} {\bibfnamefont {D.}~\bibnamefont {Chang}}, \bibinfo
  {author} {\bibfnamefont {H.}~\bibnamefont {Pichler}}, \bibinfo {author}
  {\bibfnamefont {S.~F.} \bibnamefont {Yelin}}, \bibinfo {author}
  {\bibfnamefont {P.}~\bibnamefont {Zoller}},  and \bibinfo {author}
  {\bibfnamefont {M.~D.} \bibnamefont {Lukin}}, }\bibfield  {title} {\enquote
  {\bibinfo {title} {Topological quantum optics in two-dimensional atomic
  arrays},} }\href {https://arxiv.org/abs/1703.04849} {\bibfield  {journal}
  {\bibinfo  {journal} {arXiv:1703.04849} } (\bibinfo {year}
  {2017})}\BibitemShut {NoStop}%
\bibitem [{\citenamefont {Asenjo-Garcia} \emph {et~al.}(2017)\citenamefont
  {Asenjo-Garcia}, \citenamefont {Moreno-Cardoner}, \citenamefont {Albrecht},
  \citenamefont {Kimble}, and \citenamefont {Chang}}]{asenjo2017exponential}%
  \BibitemOpen
  \bibfield  {author} {\bibinfo {author} {\bibfnamefont {A.}~\bibnamefont
  {Asenjo-Garcia}}, \bibinfo {author} {\bibfnamefont {M.}~\bibnamefont
  {Moreno-Cardoner}}, \bibinfo {author} {\bibfnamefont {A.}~\bibnamefont
  {Albrecht}}, \bibinfo {author} {\bibfnamefont {H.~J.} \bibnamefont {Kimble}},
   and \bibinfo {author} {\bibfnamefont {D.~E.} \bibnamefont {Chang}},
  }\bibfield  {title} {\enquote {\bibinfo {title} {Exponential improvement in
  photon storage fidelities using subradiance and "selective radiance" in
  atomic arrays},} }\href {https://arxiv.org/abs/1703.03382} {\bibfield
  {journal} {\bibinfo  {journal} {arXiv:1703.03382} } (\bibinfo {year}
  {2017})}\BibitemShut {NoStop}%
\bibitem [{\citenamefont {Dicke}(1954)}]{dicke1954coherence}%
  \BibitemOpen
  \bibfield  {author} {\bibinfo {author} {\bibfnamefont {R.~H.} \bibnamefont
  {Dicke}}, }\bibfield  {title} {\enquote {\bibinfo {title} {Coherence in
  spontaneous radiation processes},} }\href {\doibase 10.1103/PhysRev.93.99}
  {\bibfield  {journal} {\bibinfo  {journal} {Phys. Rev.} }\textbf {\bibinfo
  {volume} {93}}, \bibinfo {pages} {99--110} (\bibinfo {year}
  {1954})}\BibitemShut {NoStop}%
\bibitem [{\citenamefont {Gross} and \citenamefont
  {Haroche}(1982)}]{haroche1982superradiance}%
  \BibitemOpen
  \bibfield  {author} {\bibinfo {author} {\bibfnamefont {M.}~\bibnamefont
  {Gross}} and \bibinfo {author} {\bibfnamefont {S.}~\bibnamefont {Haroche}},
  }\bibfield  {title} {\enquote {\bibinfo {title} {Superradiance: {A}n essay on
  the theory of collective spontaneous emission},} }\href
  {http://www.sciencedirect.com/science/article/pii/0370157382901028}
  {\bibfield  {journal} {\bibinfo  {journal} {Physics Reports} }\textbf
  {\bibinfo {volume} {93}}, \bibinfo {pages} {301 -- 396} (\bibinfo {year}
  {1982})}\BibitemShut {NoStop}%
\bibitem [{\citenamefont {Zoubi} and \citenamefont
  {Ritsch}(2008)}]{zoubi2008bright}%
  \BibitemOpen
  \bibfield  {author} {\bibinfo {author} {\bibfnamefont {H.}~\bibnamefont
  {Zoubi}} and \bibinfo {author} {\bibfnamefont {H.}~\bibnamefont {Ritsch}},
  }\bibfield  {title} {\enquote {\bibinfo {title} {Bright and dark excitons in
  an atom-pair--filled optical lattice within a cavity},} }\href@noop {}
  {\bibfield  {journal} {\bibinfo  {journal} {EPL (Europhysics Letters)}
  }\textbf {\bibinfo {volume} {82}}, \bibinfo {pages} {14001} (\bibinfo {year}
  {2008})}\BibitemShut {NoStop}%
\bibitem [{\citenamefont {Sames} \emph {et~al.}(2014)\citenamefont {Sames},
  \citenamefont {Chibani}, \citenamefont {Hamsen}, \citenamefont {Altin},
  \citenamefont {Wilk}, and \citenamefont {Rempe}}]{sames2014antiresonance}%
  \BibitemOpen
  \bibfield  {author} {\bibinfo {author} {\bibfnamefont {C.}~\bibnamefont
  {Sames}}, \bibinfo {author} {\bibfnamefont {H.}~\bibnamefont {Chibani}},
  \bibinfo {author} {\bibfnamefont {C.}~\bibnamefont {Hamsen}}, \bibinfo
  {author} {\bibfnamefont {P.~A.} \bibnamefont {Altin}}, \bibinfo {author}
  {\bibfnamefont {T.}~\bibnamefont {Wilk}},  and \bibinfo {author}
  {\bibfnamefont {G.}~\bibnamefont {Rempe}}, }\bibfield  {title} {\enquote
  {\bibinfo {title} {Antiresonance phase shift in strongly coupled cavity
  {QED}},} }\href {\doibase 10.1103/PhysRevLett.112.043601} {\bibfield
  {journal} {\bibinfo  {journal} {Phys. Rev. Lett.} }\textbf {\bibinfo {volume}
  {112}}, \bibinfo {pages} {043601} (\bibinfo {year} {2014})}\BibitemShut
  {NoStop}%
\bibitem [{\citenamefont {Rice} and \citenamefont
  {Brecha}(1996)}]{rice1996cavity}%
  \BibitemOpen
  \bibfield  {author} {\bibinfo {author} {\bibfnamefont {P.~R.} \bibnamefont
  {Rice}} and \bibinfo {author} {\bibfnamefont {R.~J.} \bibnamefont {Brecha}},
  }\bibfield  {title} {\enquote {\bibinfo {title} {Cavity induced
  transparency},} }\href
  {http://www.sciencedirect.com/science/article/pii/0030401896001022}
  {\bibfield  {journal} {\bibinfo  {journal} {Optics Communications} }\textbf
  {\bibinfo {volume} {126}}, \bibinfo {pages} {230 -- 235} (\bibinfo {year}
  {1996})}\BibitemShut {NoStop}%
\bibitem [{\citenamefont {Lehmberg}(1970)}]{lehmberg1970radiation}%
  \BibitemOpen
  \bibfield  {author} {\bibinfo {author} {\bibfnamefont {R.}~\bibnamefont
  {Lehmberg}}, }\bibfield  {title} {\enquote {\bibinfo {title} {Radiation from
  an {N}-atom system. {I}. {G}eneral formalism},} }\href {\doibase
  https://doi.org/10.1103/PhysRevA.2.883} {\bibfield  {journal} {\bibinfo
  {journal} {Phys. Rev. A} }\textbf {\bibinfo {volume} {2}}, \bibinfo {pages}
  {883} (\bibinfo {year} {1970})}\BibitemShut {NoStop}%
\bibitem [{\citenamefont {Gardiner} and \citenamefont
  {Zoller}(2004)}]{gardiner2004quantum}%
  \BibitemOpen
  \bibfield  {author} {\bibinfo {author} {\bibfnamefont {C.}~\bibnamefont
  {Gardiner}} and \bibinfo {author} {\bibfnamefont {P.}~\bibnamefont {Zoller}},
  }\href@noop {} {\emph {\bibinfo {title} {Quantum noise: a handbook of
  Markovian and non-Markovian quantum stochastic methods with applications to
  quantum optics}}}, Vol.~\bibinfo {volume} {56} (\bibinfo  {publisher}
  {Springer Science \& Business Media}, \bibinfo {year} {2004})\BibitemShut
  {NoStop}%
\bibitem [{\citenamefont {Alsing} \emph {et~al.}(1992)\citenamefont {Alsing},
  \citenamefont {Cardimona}, and \citenamefont
  {Carmichael}}]{alsing1992suppression}%
  \BibitemOpen
  \bibfield  {author} {\bibinfo {author} {\bibfnamefont {P.~M.} \bibnamefont
  {Alsing}}, \bibinfo {author} {\bibfnamefont {D.~A.} \bibnamefont
  {Cardimona}},  and \bibinfo {author} {\bibfnamefont {H.~J.} \bibnamefont
  {Carmichael}}, }\bibfield  {title} {\enquote {\bibinfo {title} {Suppression
  of fluorescence in a lossless cavity},} }\href {\doibase
  10.1103/PhysRevA.45.1793} {\bibfield  {journal} {\bibinfo  {journal} {Phys.
  Rev. A} }\textbf {\bibinfo {volume} {45}}, \bibinfo {pages} {1793--1803}
  (\bibinfo {year} {1992})}\BibitemShut {NoStop}%
\bibitem [{\citenamefont {Zippilli} \emph {et~al.}(2004)\citenamefont
  {Zippilli}, \citenamefont {Morigi}, and \citenamefont
  {Ritsch}}]{zippilli2004suppression}%
  \BibitemOpen
  \bibfield  {author} {\bibinfo {author} {\bibfnamefont {S.}~\bibnamefont
  {Zippilli}}, \bibinfo {author} {\bibfnamefont {G.}~\bibnamefont {Morigi}},
  and \bibinfo {author} {\bibfnamefont {H.}~\bibnamefont {Ritsch}}, }\bibfield
  {title} {\enquote {\bibinfo {title} {Suppression of {B}ragg scattering by
  collective interference of spatially ordered atoms with a high-q cavity
  mode},} }\href@noop {} {\bibfield  {journal} {\bibinfo  {journal} {Phys. Rev.
  Lett.} }\textbf {\bibinfo {volume} {93}}, \bibinfo {pages} {123002} (\bibinfo
  {year} {2004})}\BibitemShut {NoStop}%
\bibitem [{\citenamefont {Ostermann} \emph {et~al.}(2014)\citenamefont
  {Ostermann}, \citenamefont {Plankensteiner}, \citenamefont {Ritsch}, and
  \citenamefont {Genes}}]{ostermann2014protected}%
  \BibitemOpen
  \bibfield  {author} {\bibinfo {author} {\bibfnamefont {L.}~\bibnamefont
  {Ostermann}}, \bibinfo {author} {\bibfnamefont {D.}~\bibnamefont
  {Plankensteiner}}, \bibinfo {author} {\bibfnamefont {H.}~\bibnamefont
  {Ritsch}},  and \bibinfo {author} {\bibfnamefont {C.}~\bibnamefont {Genes}},
  }\bibfield  {title} {\enquote {\bibinfo {title} {Protected subspace {R}amsey
  spectroscopy},} }\href {\doibase 10.1103/PhysRevA.90.053823} {\bibfield
  {journal} {\bibinfo  {journal} {Phys. Rev. A} }\textbf {\bibinfo {volume}
  {90}}, \bibinfo {pages} {053823} (\bibinfo {year} {2014})}\BibitemShut
  {NoStop}%
\bibitem [{\citenamefont {Mascarenhas} \emph {et~al.}(2013)\citenamefont
  {Mascarenhas}, \citenamefont {Gerace}, \citenamefont {Santos}, and
  \citenamefont {Auff\`eves}}]{mascarenhas2013cooperativity}%
  \BibitemOpen
  \bibfield  {author} {\bibinfo {author} {\bibfnamefont {E.}~\bibnamefont
  {Mascarenhas}}, \bibinfo {author} {\bibfnamefont {D.}~\bibnamefont {Gerace}},
  \bibinfo {author} {\bibfnamefont {M.~F.} \bibnamefont {Santos}},  and
  \bibinfo {author} {\bibfnamefont {A.}~\bibnamefont {Auff\`eves}}, }\bibfield
  {title} {\enquote {\bibinfo {title} {Cooperativity of a few quantum emitters
  in a single-mode cavity},} }\href {\doibase 10.1103/PhysRevA.88.063825}
  {\bibfield  {journal} {\bibinfo  {journal} {Phys. Rev. A} }\textbf {\bibinfo
  {volume} {88}}, \bibinfo {pages} {063825} (\bibinfo {year}
  {2013})}\BibitemShut {NoStop}%
\bibitem [{\citenamefont {Jenkins} and \citenamefont
  {Ruostekoski}(2012)}]{jenkins2012controlled}%
  \BibitemOpen
  \bibfield  {author} {\bibinfo {author} {\bibfnamefont {S.~D.} \bibnamefont
  {Jenkins}} and \bibinfo {author} {\bibfnamefont {J.}~\bibnamefont
  {Ruostekoski}}, }\bibfield  {title} {\enquote {\bibinfo {title} {Controlled
  manipulation of light by cooperative response of atoms in an optical
  lattice},} }\href {\doibase 10.1103/PhysRevA.86.031602} {\bibfield  {journal}
  {\bibinfo  {journal} {Phys. Rev. A} }\textbf {\bibinfo {volume} {86}},
  \bibinfo {pages} {031602} (\bibinfo {year} {2012})}\BibitemShut {NoStop}%
\bibitem [{\citenamefont {Jenkins} and \citenamefont
  {Ruostekoski}(2013)}]{jenkins2013metamaterial}%
  \BibitemOpen
  \bibfield  {author} {\bibinfo {author} {\bibfnamefont {S.~D.} \bibnamefont
  {Jenkins}} and \bibinfo {author} {\bibfnamefont {J.}~\bibnamefont
  {Ruostekoski}}, }\bibfield  {title} {\enquote {\bibinfo {title} {Metamaterial
  transparency induced by cooperative electromagnetic interactions},} }\href
  {\doibase 10.1103/PhysRevLett.111.147401} {\bibfield  {journal} {\bibinfo
  {journal} {Phys. Rev. Lett.} }\textbf {\bibinfo {volume} {111}}, \bibinfo
  {pages} {147401} (\bibinfo {year} {2013})}\BibitemShut {NoStop}%
\bibitem [{\citenamefont {Juan} \emph {et~al.}(2016)\citenamefont {Juan},
  \citenamefont {Molina-Terriza}, \citenamefont {Volz}, and \citenamefont
  {Romero-Isart}}]{juan2016near}%
  \BibitemOpen
  \bibfield  {author} {\bibinfo {author} {\bibfnamefont {M.~L.} \bibnamefont
  {Juan}}, \bibinfo {author} {\bibfnamefont {G.}~\bibnamefont
  {Molina-Terriza}}, \bibinfo {author} {\bibfnamefont {T.}~\bibnamefont
  {Volz}},  and \bibinfo {author} {\bibfnamefont {O.}~\bibnamefont
  {Romero-Isart}}, }\bibfield  {title} {\enquote {\bibinfo {title} {Near-field
  levitated quantum optomechanics with nanodiamonds},} }\href {\doibase
  10.1103/PhysRevA.94.023841} {\bibfield  {journal} {\bibinfo  {journal} {Phys.
  Rev. A} }\textbf {\bibinfo {volume} {94}}, \bibinfo {pages} {023841}
  (\bibinfo {year} {2016})}\BibitemShut {NoStop}%
\bibitem [{\citenamefont {Dantan} \emph {et~al.}(2014)\citenamefont {Dantan},
  \citenamefont {Nair}, \citenamefont {Pupillo}, and \citenamefont
  {Genes}}]{dantan2014hybrid}%
  \BibitemOpen
  \bibfield  {author} {\bibinfo {author} {\bibfnamefont {A.}~\bibnamefont
  {Dantan}}, \bibinfo {author} {\bibfnamefont {B.}~\bibnamefont {Nair}},
  \bibinfo {author} {\bibfnamefont {G.}~\bibnamefont {Pupillo}},  and \bibinfo
  {author} {\bibfnamefont {C.}~\bibnamefont {Genes}}, }\bibfield  {title}
  {\enquote {\bibinfo {title} {Hybrid cavity mechanics with doped systems},}
  }\href {journals.aps.org/pra/abstract/10.1103/PhysRevA.90.033820} {\bibfield
  {journal} {\bibinfo  {journal} {Phys. Rev. A} }\textbf {\bibinfo {volume}
  {90}}, \bibinfo {pages} {033820} (\bibinfo {year} {2014})}\BibitemShut
  {NoStop}%
\bibitem [{\citenamefont {Jenkins} \emph {et~al.}(2016)\citenamefont {Jenkins},
  \citenamefont {Ruostekoski}, \citenamefont {Papasimakis}, \citenamefont
  {Savo}, and \citenamefont {Zheludev}}]{jenkins2016many}%
  \BibitemOpen
  \bibfield  {author} {\bibinfo {author} {\bibfnamefont {S.~D.} \bibnamefont
  {Jenkins}}, \bibinfo {author} {\bibfnamefont {J.}~\bibnamefont
  {Ruostekoski}}, \bibinfo {author} {\bibfnamefont {N.}~\bibnamefont
  {Papasimakis}}, \bibinfo {author} {\bibfnamefont {S.}~\bibnamefont {Savo}},
  and \bibinfo {author} {\bibfnamefont {N.~I.} \bibnamefont {Zheludev}},
  }\bibfield  {title} {\enquote {\bibinfo {title} {Many-body subradiant
  excitations in metamaterial arrays: Experiment and theory},} }\href
  {https://arxiv.org/abs/1611.01509} {\bibfield  {journal} {\bibinfo  {journal}
  {arXiv preprint arXiv:1611.01509} } (\bibinfo {year} {2016})}\BibitemShut
  {NoStop}%
\bibitem [{\citenamefont {Hunter}(2007)}]{hunter2007matplotlib}%
  \BibitemOpen
  \bibfield  {author} {\bibinfo {author} {\bibfnamefont {J.~D.} \bibnamefont
  {Hunter}}, }\bibfield  {title} {\enquote {\bibinfo {title} {Matplotlib: {A}
  2{D} graphics environment},} }\href {\doibase 10.1109/MCSE.2007.55}
  {\bibfield  {journal} {\bibinfo  {journal} {Comput. Sci. Eng.} }\textbf
  {\bibinfo {volume} {9}}, \bibinfo {pages} {90--95} (\bibinfo {year}
  {2007})}\BibitemShut {NoStop}%
\end{thebibliography}%

\clearpage

\renewcommand{\theequation}{A.\arabic{equation}}
\renewcommand\thefigure{A.\arabic{figure}}
\renewcommand{\section}[1]{\vspace{.1cm}\emph{Appendix: #1} --}
\setcounter{equation}{0}
\setcounter{figure}{0}

\appendix

\section{Vacuum mediated coherent and incoherent dynamics}
For a collection of emitters at positions $\textbf{r}_i$ the collective spontaneous emission rates and the coherent dipole-dipole interaction strengths are \cite{lehmberg1970radiation}
\begin{align}
\gamma_{ij} &= \frac{3\gamma}{2}F(\textbf{k}\ts{e}\cdot\textbf{r}_{ij}),
\\
\Omega_{ij} &= -\frac{3\gamma}{4}G(\textbf{k}\ts{e}\cdot\textbf{r}_{ij}),
\end{align}
respectively. Here, $\gamma$ is the single emitter free space decay rate, $\textbf{k}\ts{e}$ is the transition wave vector and $\textbf{r}_{ij}:=\textbf{r}_i-\textbf{r}_j$. The functions $F$ and $G$ are defined as
\begin{align}
F(x) &= \left(1-\cos^2\theta\right)\frac{\sin x}{x} +
\notag \\
&+ \left(1-3\cos^2\theta\right)\left(\frac{\cos x}{x^2}-\frac{\sin x}{x^3}\right),
\\
G(x) &= -\left(1-\cos^2\theta\right)\frac{\cos x}{x} +
\notag \\
&+ \left(1-3\cos^2\theta\right)\left(\frac{\sin x}{x^2} + \frac{\sin x}{x^3}\right),
\end{align}
where $\theta$ is the angle drawn by the dipole moment $\boldsymbol{\mu}$ and the separation vector $\textbf{r}_{ij}$. Note, that for all computations in the paper we assumed the dipoles to be oriented along the y-direction.

\section{Cavity input-output relations}
Consider a cavity with two mirrors $A$ and $B$. We drive the cavity through the mirror $B$ and measure the output at the opposite mirror $A$. Furthermore, we assume that both mirrors have identical losses of $\kappa/2$. The input-output relations of the total input and output operators (i.e. the input white noise on top of the classical input) for both mirrors are
\begin{align}
B_\text{in} + B_\text{out} &= \sqrt{\kappa}a,
\\
A_\text{in} + A_\text{out} &= \sqrt{\kappa}a.
\end{align}
Taking the classical average of the above equations, and assuming the drive through $B$ at amplitude $\braket{B_\text{in}}=\eta/\sqrt{\kappa}$,
\begin{align}
\frac{\eta}{\sqrt{\kappa}} + \braket{B_\text{out}} &= \sqrt{\kappa}\braket{a},
\\
\braket{A_\text{out}} &= \sqrt{\kappa}\braket{a}.
\label{A_out_relation}
\end{align}
Hence, we find that the output at port $A$ reflects the cavity field according to \eqref{A_out_relation}.

\section{Emitter input noise in the QLE}
Writing the input noise in the QLEs for the emitter ensemble is not straightforward as the decay is not diagonal. We first note that for a diagonal Lindblad term describing the damping of a system via multiple damping channels with operators $c_j$ and rates $\nu_j$, the QLE~\cite{gardiner2004quantum} of an arbitrary system operator $A$ is
\begin{align} \label{diagonal_HLE}
\dot{A} &= i[H,A] - \sum_j[A, c_j^\dag]\left(\frac{\nu_j}{2}c + \sqrt{\nu_j}c_\text{in}\right) +
\notag \\
&+ \sum_j\left(\frac{\nu_j}{2}c^\dag + \sqrt{\nu_j}c_\text{in}^\dag\right)[A,c],
\end{align}
where $c_\text{in}$ is the uncorrelated white input noise $\braket{c_\text{in}(t)c_\text{in}^\dag(t')}=\delta(t-t')$.

For correlated emitters, it is possible to diagonalize the Lindblad term such that each decay channel is described by a damping operator which is a linear combination of all $\sigma_j^-$. We may perform the transformation back to the original non-diagonal form in order to find the input noise terms. To this end, let \textbf{T} be the real and orthogonal ($\textbf{T}^{-1}=\textbf{T}^\text{T}$) matrix which diagonalizes the matrix $\boldsymbol{\Gamma}$ as
\begin{align}
\text{diag}\left(\lambda_1,...,\lambda_N\right) &= \textbf{T}^{-1}\boldsymbol{\Gamma}\textbf{T},
\end{align}
where $\lambda_j$ is the $j$th eigenvalue of the decay matrix. Defining a set of damping operators
\begin{align}\label{Pi_op}
\Pi_j^\pm &:= \sum_k \left(\textbf{T}^{-1}\right)_{jk}\sigma_k^\pm,
\end{align}
we may write \cite{ostermann2014protected}
\begin{align}
\mathcal{L}_\text{e}[\rho] &= \sum_i\frac{\lambda_i}{2}\left(2\Pi_i^-\rho\Pi_i^+ - \Pi_i^+\Pi_i^-\rho - \rho\Pi_i^+\Pi_i^-\right).
\end{align}
Obviously, this Lindblad term is diagonal and hence the QLE may be cast into the form given by \eqref{diagonal_HLE}. The input noise terms of the emitter operators $\sigma_{i,\text{in}}^\pm$ follow the transformation rules given by \eqref{Pi_op}. Transforming the QLE for any emitter operator $A$ back into the non-diagonal form gives the usual terms for the deterministic parts. For the noise terms, however, we have
\begin{align}
\sum_j[A,\Pi_j^+]\sqrt{\lambda_j}\Pi_{j,\text{in}}^- &= \sum_j[A,\sigma_j^+]\xi_j(t),
\\
\sum_j[A,\Pi_j^-]\sqrt{\lambda_j}\Pi_{j,\text{in}}^+ &= \sum_j[A,\sigma_j^-]\xi_j^\dag(t),
\end{align}
where we have implicitly defined our correlated emitter noise terms $\xi_j$ as
\begin{align}
\xi_j(t) &:= \sum_{k,l} \textbf{T}_{jk}\sqrt{\lambda_k}(\textbf{T}^{-1})_{kl}\sigma_{l,\text{in}}^-.
\end{align}
Hence, the QLE for any emitter operator $A$ is
\begin{align}
\dot{A} &= i[H,A] - \sum_{ij}[A,\sigma_i^+]\left(\frac{\gamma_{ij}}{2}\sigma_j^- + \xi_i(t)\right) +
\notag \\
&+ \sum_{ij} \left(\frac{\gamma_{ij}}{2}\sigma_j^+ + \xi_i^\dag(t)\right)[A,\sigma_i^-],
\end{align}
with the spatially correlated white noise $\xi_i$. From the definition of the noise it is straightforward to show that
\begin{align}
\braket{\xi_i(t)\xi_j^\dag(t')} &= \gamma_{ij}\delta(t-t').
\end{align}

\section{Lorentzian shape of the antiresonance}
We verify the Lorentzian profile of the antiresonance for a single emitter interacting with a single cavity mode by fitting
\begin{align}
B(\Delta)=\left|\frac{\kappa}{i\Delta+\kappa}\right|^2-\left|\frac{\kappa}{i\Delta+\kappa+g^2/(i\Delta + \gamma)}\right|^2,
\end{align}
with a Lorentzian
\begin{align}
\left|\frac{\beta s}{i\Delta+\beta}\right|^2,
\end{align}
where $\beta$ gives the linewidth while $s$ gives the height. We compute immediately
\begin{align}
s=1-\frac{1}{(1+C)^2},
\end{align}
and from
\begin{align}
\beta &= \sqrt{\frac{2B(0)}{\frac{\partial^2 B}{\partial\Delta^2}(0)}}
\end{align}
we obtain
\begin{widetext}
\begin{align}
\beta = \sqrt{\frac{\kappa^2(g^2+\kappa \gamma)^2(g^2+2\kappa \gamma)}{g^6+4g^4 \kappa \gamma + 2 \kappa^3 \gamma (\kappa^2 +\kappa \gamma +2 \gamma^2)+g^2(\kappa^4 +6 \kappa^2 \gamma^2)}}.
\end{align}
\end{widetext}
In the limit that both $\gamma,g\ll\kappa$ when only keeping terms of $\mathcal{O}(\kappa^4)$ and higher, this reduces to $\beta\approx\gamma(1+C)$.

The results obtained above are valid for two emitters with the replacements $\gamma\to\gamma\ts{eff}$, $C\to C\ts{eff}$ and the detuning $\Delta\ts{c}=\Delta\ts{eff}=\Delta$.

\section{Characterizing the phase of the field}

In order to describe the phase analytically, consider that for the single emitter transmission coefficient with $\Delta\ts{c}=\Delta\ts{e}=\Delta$, we have
\begin{align} 
\frac{\Im\{t\}}{\Re\{t\}} &= \frac{\Delta\left(g^2 - \Delta^2 - \gamma^2\right)}{\kappa\left(\gamma^2 + \Delta^2\right) + \gamma g^2}.
\end{align}
The phase of the field is then given by the arctangent of the above expression. For sufficiently small phase shifts ($C\ll 1$), the phase is well approximated by the argument of the arctangent, such that
\begin{align} \label{phi_app}
\frac{\partial}{\partial\Delta}\left(\phi - \phi\ts{c}\right) &\approx \frac{C\gamma\left(\kappa + \gamma\right)\left(\gamma^2\left(1 + C\right) - \Delta^2\right)}{\kappa\left(\gamma^2 \left(1 + C\right) + \Delta^2\right)^2}.
\end{align}
Finding the roots of this expression yields the detuning at which the phase shift of the emitter is maximal (positive $\Delta$) or minimal (negative $\Delta$), respectively,
\begin{align}
\Delta &= \pm\gamma\sqrt{1 + C}.
\end{align}
We note, that for large cooperativity the approximation of the phase does no longer hold, however, given that $g,\gamma\ll\kappa$ the position of the maximum and minimum are well approximated by the above detuning even if $C\gg 1$, such that
\begin{align}
\max_\Delta\left[\phi - \phi\ts{c}\right] &\approx \max_\Delta\left[\phi\right] \approx \arctan\left(\frac{C}{2\sqrt{1 + C}}\right).
\end{align}
Here we wrote down the maximum of the phase and note that the minimum only differs from this expression with a negative sign. Now, depending on the sign of $\Delta$, taking the limit of $\gamma\to 0$, the phase either goes to $\pi/2$ or $-\pi/2$, while $\Delta\to0$, i.e.
\begin{align}
\lim_{\Delta\to0^\pm} \phi = \pm\frac{\pi}{2}.
\end{align}
This means that the phase exhibits a jump of magnitude $\pi$ when crossing the resonance, which can also be seen when computing the slope. To this end, consider the fact that close to the resonance where $|\Delta|\ll\kappa$, the phase is small such that the approximation in Eq.~\eqref{phi_app} is valid again. In the limit of $|\Delta|\ll\gamma,g\ll\kappa$ but large cooperativity $g^2\gg \kappa\gamma$ and hence $g\gg\gamma$, this expression becomes
\begin{align}
\phi - \phi\ts{c} &\approx \phi \approx \frac{\Delta}{\gamma}.
\end{align}
We conclude that the slope at the resonance (where the phase changes sign) is proportional to $1/\gamma$ and hence diverges for $\gamma\to 0$, which is in agreement with the previous result where the phase shift is maximal or minimal at $\Delta\to 0$ depending on whether one takes the left-sided or right-sided limit.

Finally, we note that the above discussion is also applicable to two emitters when replacing $\gamma \rightarrow \gamma\ts{eff}$, $C \rightarrow C\ts{eff}$, and setting $\Delta\ts{c}=\Delta\ts{eff}=\Delta$.

\section{Matching frequency and symmetry of antiresonances}

For \fref{fig2}(e),(f) we set
\begin{align}
\textbf{G}\approx &(0.72, -1.44,  2.03, -2.46,  2.68,
\notag \\
&-2.68,  2.46, -2.03,  1.44, -0.72)^\intercal \times 10^{-2} \kappa,
\end{align}
which corresponds to the coefficients of the eigenvector of $\boldsymbol{\Omega}$ that had the largest overlap with the eigenvector of $\boldsymbol{\Gamma}$ corresponding to the smallest eigenvalue and therefore the smallest decay rate \cite{ostermann2014protected}. The frequency was matched by numerically solving $\Delta\ts{eff}(\delta)=0$, which lead to $\delta\approx 0.234\kappa$, and setting $\Delta\ts{c}=\Delta\ts{e}-\delta$.
\vfill

\section{Comparison to exact numerics}

\begin{figure}[hb]
\includegraphics[width=.8\columnwidth]{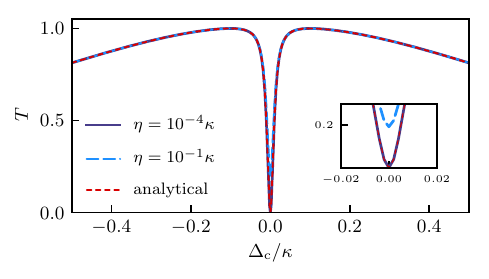}
\caption{\emph{Comparison of analytics with exact numerics}. We compare the full numerics, i.e. without using $\braket{\sigma_i^z}\approx -1$, to the analytical result in an extremely subradiant regime for a 4 emitter chain with opposite coupling and without dipole-dipole shifts ($\Omega_{ij}=0$). The inset is zoomed in on the antiresonance to emphasize the agreement. The parameters are $d=0.02\lambda\ts{e}$, $g=\kappa/20=2\gamma$ and $\Delta\ts{e}=0$.}
\label{figA1}
\end{figure}

Let us now comment on the accuracy of the linearization used to obtain the form of eqs. \eqref{single_HE1},\eqref{single_HE2} and \eqref{HE1},\eqref{HE2}. From \fref{figA1}, it is clear that as long as we keep the driving strength $\eta$ weak enough, the requirements for the low-excitation limit are fulfilled rendering the analytics exact (cf. red dots and dark blue line in \fref{figA1}). If the driving becomes too strong, the excitation of the emitters is no longer negligible resulting in a discrepancy between the full numerics and the analytics (cf. dashed, light blue line in \fref{figA1}). Nevertheless, this does not change the results qualitatively, i.e. there still is a subradiant antiresonance.

\vfill

\end{document}